\documentclass{aa}  

\usepackage{graphicx}
\usepackage{txfonts}

\bibpunct{(}{)}{;}{a}{}{,} 

\begin{document}

   \title{Dusty disk winds at the sublimation rim of the highly inclined, low mass YSO SU\,Aurigae}


   \author{Aaron Labdon
          \inst{1}
          \and
          Stefan Kraus
          \inst{1}
          \and  
          Claire L.\ Davies
          \inst{1}
          \and
          Alexander Kreplin
          \inst{1}
          \and
          Jacques Kluska
          \inst{1}
          \and
          Tim J.\ Harries
          \inst{1}
          \and
          John D.\ Monnier
          \inst{2}
          \and
          Theo ten Brummelaar
          \inst{3}
          \and
          Fabien Baron
          \inst{4}
          \and
          Rafael Millan-Gabet
          \inst{5}
          \and
          Brian Kloppenborg
          \inst{4}
          \and
          Joshua Eisner
          \inst{6}
          \and
          Judit Sturmann
          \inst{3}
          \and
          Laszlo Sturmann
          \inst{3}
          }

   \institute{
   (1) University of Exeter, School of Physics and Astronomy, Astrophysics Group, Stocker Road, Exeter, EX4 4QL, UK\\
   (2) University of Michigan, Department of Astronomy, S University Ave, Ann Arbor, MI 48109, USA\\
   (3) The CHARA Array of Georgia State University, Mount Wilson Observatory, Mount Wilson, CA 91023, USA\\
   (4) Department of Physics and Astronomy, Georgia State University, Atlanta, GA, USA\\
   (5) Infrared Processing and Analysis Center, California Institute of Technology, Pasadena, CA, 91125, USA \\
   (6) Steward Observatory, University of Arizona, Tucson, AZ, 85721, USA
   }

   \date{accepted May 18, 2019}

 
  \abstract
   {T\,Tauri stars are low-mass young stars whose disks provide the setting for planet formation. Despite this, their structure is poorly understood. We present new infrared interferometric observations of the SU Aurigae circumstellar environment that offer $3$ times higher resolution and better baseline position angle coverage over previous observations.}
   {We aim to investigate the characteristics of the circumstellar material around SU\,Aur, constrain the disk geometry, composition and inner dust rim structure.
   }
   {The CHARA array offers unique opportunities for long baseline observations, with baselines up to $331$\,m. Using the CLIMB 3-telescope combiner in the K-band allows us to measure visibilities as well as closure phase.
   We undertook image reconstruction for model-independent analysis, and fitted geometric models such as Gaussian and ring distributions. 
   Additionally, the fitting of radiative transfer models constrain the physical parameters of the disk. 
   For the first time, a dusty disk wind is introduced to the radiative transfer code TORUS to model protoplanetary disks.
   Our implementation is motivated by theoretical models of dusty disk winds, where magnetic field lines drive dust above the disk plane close to the sublimation zone.
   }
   {
   Image reconstruction reveals an inclined disk with slight asymmetry along its minor-axis, likely due to inclination effects obscuring the inner disk rim through absorption of incident star light on the near-side and thermal re-emission/scattering of the far-side. 
   Geometric modelling of a skewed ring finds the inner rim at $0.17\pm0.02~\mathrm{au}$ with an inclination of $50.9\pm1.0^\circ$ and minor axis position angle $60.8\pm1.2^\circ$. 
   Radiative transfer modelling shows a flared disk with an inner radius at $0.18$\,au which implies a grain size of $0.4\,\mu$m assuming astronomical silicates and a scale height of $15.0$\,au at $100$\,au. 
   Among the tested radiative transfer models, only the dusty disk wind successfully accounts for the K-band excess by introducing dust above the mid-plane.
   }
   {}


   \maketitle
%

\section{Introduction} \label{sec:intro}
    Protoplanetary disks are observed across all masses of young stellar objects \citep{Lazareff17,Kraus17b}. Created as a consequence of the conservation of angular momentum during the star formation process, infalling gas and dust form a circumstellar disk around protostellar objects. Circumstellar disks are understood to be the birthplace of planets. The scattering and thermal re-emission of starlight by optically thick dust in the inner disk induces stong near-infrared (NIR) excess emission.

    \begin{table}[b!]
        \caption{\label{table:Stellar}Stellar parameters of SU\,Aurigae}
        \centering
        \begin{tabular}{c c c} 
            \hline
            \noalign{\smallskip}
            Parameter  &  Value &   Reference   \\ [0.5ex]
            \hline
            \noalign{\smallskip}
            RA (J2000) &  $04~55~59.39$   & (1)  \\
            DEC (J2000) & $+30~34~01.50$  & (1)          \\
            Mass & $2.0~M_\mathrm{\odot}$ & (2) \\ 
            Sp. Type  & G2~IIIe & (2)  \\
            Distance & $157.68^{+1.49}_{-1.48}~\mathrm{pc}$ & (1)  \\
            Radius & $3.5~R_\mathrm{\odot}$ & (2) \\
            $K_{mag} $ & $5.99$ & (3) \\
            $T_\mathrm{{eff}}$ & $5860~\mathrm{K}$ & (2)  \\ [1ex] 
            \hline
        \end{tabular}
        \tablebib{
        (1) \citet{Gaia}; (2)~\citet{DeWarf}; (3)~\citet{Cutri03}
        }
    \end{table}
    
    Developments in the field of optical interferometry allowed for the first spatially resolved observations inner astronomical unit-scale of young stellar objects (YSOs) \citep{Millan-Gabet99,Akeson00}. It is within these inner regions that important processes, such as star-disk interactions (accretion and outflows), dust sublimation and planet formation occur. Accretion processes are one of the dominant effects in planet formation in the inner disk. Internal friction, or viscosity within the disk drives accretion onto the central star. In order to preserve angular momentum, some material is lost through outflow processes such as jets \citep{Williams11}. Similarly, outflowing material such as magnetically driven winds can remove angular momentum from the disk, efficiently driving accretion processes \citep{Turner14}. The current generation of interferometers provide unparalleled opportunities to study the smallest scales of disk structures \citep{Davies18,Setterholm18}, in particular the Centre for High Angular Resolution Astronomy (CHARA) can provide baselines up to $331$\,m allowing us to study the very inner disk regions.
    
    The first inner rim models of protoplanetary disks based upon spectral energy distribution (SED) analysis adopted a vertical inner wall of emission \citep{Dullemond01,Natta01}. However, interferometric observations failed to find evidence of the strong asymmetries predicted at high disk inclinations by this model \citep{monnier06,Kraus2009B}. An alternative curved rim model was later proposed, with a curvature arising from a gas-density-dependent sublimation temperature \citep{Isella05}. The brightness distribution predicted by these models is much more symmetric in nature. However, these models predict a very pronounced second visibility lobe, which has yet to be observed in other protoplanetary disks such as HD\,142666 \citep{Davies18}, AB\,Aur, V\,1295\,Aql and MWC\,275 \citep{Tannirkulam08,Setterholm18}. Instead a much flatter second lobe is observed, indicating that a significant fraction of NIR flux originates outside of standard disk models. These rim-only models are as such insufficient to explain many interferometric observations.

    Several physical mechanisms have been proposed to explain the observed emission outside of the dust sublimation rim. One prominent explanation is the existence of dusty disk winds which send material along magnetic fields near the dust sublimation zone. This allows for optically thick material to exist close enough to the central star to contribute to the NIR emission above the disk structure. This model has been shown to successfully account for the NIR excess of the SED and the basic visibility features of AB\,Aur, MWC\,275 and RY\,Tau \citep{Konigl11,Petrov19}.
    
     \begin{figure}[t!]
        \centering
        \includegraphics[scale=0.7]{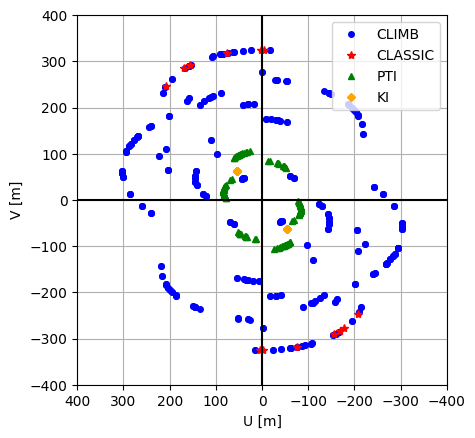}
        \caption{Coverage of the uv plane of the interferometric observations. Blue points represent observations  using the CLIMB instrument, while red stars represent the CLASSIC instrument at the CHARA array. Yellow triangles represent observations using the Keck interferometer in 2011 and green triangles using the PTI with dates between 1999 and 2004.}
        \label{fig:uvplane}
    \end{figure}
    
    \object{SU\,Aurigae} (SU\,Aur) is a $5.18\pm0.13\,\mathrm{Myr}$ old \citep{Bochanski18} low mass pre-main-sequence star in the Upper Sco star forming region at a distance of $158^{+1.49}_{-1.48}\,\mathrm{pc}$, obtained from GAIA DR2 parallax measurements \citep{GAIA2par}. As a G2-type star it has a similar effective temperature to the sun \citep{DeWarf}, but a much higher bolometric luminosity at $12.06\,\mathrm{L_\odot}$ \citep[calculated from GAIA DR2,][]{GAIA2phot} putting it in the sub-giant star class. The stellar parameters adopted are listed in Table~\ref{table:Stellar}. SU\,Aur is known to be variable in the V\,band, varying up to $0.5$\,mag over a several day cycle \citep{Unruh04}. On the other hand variability in the K\,band is minimal \citep{Akeson05} allowing us to assume the flux contribution from the star is constant across different epochs of observations. As such, any variation in the visibility is likely geometric. 
    
    Spectroscopic and photometric monitoring of SU\,Aur by \citet{Petrov19} has revealed that a dusty disk wind is the potential source of the photometric variability in both SU\,Aur and RY\,Tau. The characteristic time of change in the disk wind outflow velocity and the stellar brightness indicate that the obscuring dust is located close to the sublimation rim of the disk, in agreement with previous theoretical disk wind models \citep{Bans12,Konigl11}. 
    
    Interferometric observations carried out by \citet{Akeson05} using the Palomar Testbed Interferometer (PTI), a 3-telescope interferometer with baselines up to $110$\,m, found a disk inclined at $62^{\circ+4}_{-8}$ with a minor axis position angle of $24\pm23^\circ$ and a sublimation rim at $0.21$\,au. The disk was modelled as a flared disk with a vertical inner wall of emission using radiative transfer to fit both interferometric and photometric data. However, optically thick gas close to the central star was needed to fit the SED. The disk geometry is in agreement with values from \citet{Eisner14} using the Keck interferometer with values derived from an upper limit to the fit of the Br$\gamma$ emission. They find an upper limit on the disk inclination of $50^\circ$ and minor axis position angle of $50^\circ$. The difference in the derived position angles in previous studies is likely due to the lack of Br$\gamma$ emission and poor signal to noise of \citet{Eisner14} and the limited baseline range of \citet{Akeson05}. Hence, new observations with significantly better uv coverage on much longer baselines were needed. Additionally, polarimetric imaging campaigns, undertaken by \citet{deLeon15} and \citet{Jeffers14} revealed the presence of tails protruding from the disk at both H-band and visible wavelengths, likely associated with an extended reflection nebula and a possible undetected brown dwarf companion encounter. A companion search was undertaken by the SEEDS (Strategic Exploration of Exoplanets and Disks with Subaru) imaging survey and ruled out the presence of a companion down to $10$ Jupiter masses at separations down to $15\,\mathrm{au}$, contradicting the potential brown dwarf encounter theory of \citet{deLeon15}.     
    
    This paper presents the lowest mass YSO to be studied with very long baseline ($>110$\,m) NIR interferometry to date and is the first study probing the detailed rim structure of SU\,Aur with interferometric observation on baselines up to $331$\,m. Three different modelling methodologies were applied: (i) Image reconstruction was used to obtain a model-independent representation of the data and to derive the basic object morphology. (ii) Following this geometric model fitting allowed us to gain an appreciation for the viewing geometry of the disk by fitting Gaussian and ring models to the data. (iii) Finally, we combine interferometry and photometry to derive physical parameters with radiative transfer analysis, where our particular focus is on the physical characteristics of the inner rim.

    \begin{table*}[ht]
        \caption{\label{table:LOG}Observing log from 1999 to 2014 from the CHARA, KI and PTI interferometers.}
        
        \centering
        \begin{tabular}{c c c c c } 
            \hline
            \noalign{\smallskip}
            Date  &  Beam Combiner &   Stations  & Pointings & Calibrator (UD [mas]) \\ [0.5ex]
            \hline
            \noalign{\smallskip}
            2010-10-02 &  CHARA/CLIMB   & S1-E1-W1   & 2 & \object{HD 29867} ($0.280\pm0.007$), \object{HD 34499} ($0.257\pm0.006$)\\
            2010-12-02 &  CHARA/CLIMB   & S2-E1-W2  & 1 & \object{HD 32480} ($0.236\pm0.006$)\\
            2010-12-03 &  CHARA/CLIMB   & (S2)-E1-W2  & 1 & \object{HD 32480} ($0.236\pm0.006$), \object{HD 36724} ($0.233\pm0.006$)\\
            2012-10-18 &  CHARA/CLIMB   & S1-E1-W1  & 2 & \object{HD 27777} ($0.204\pm0.006$), 34053\tablefootmark{*}\\
            2012-10-19 &  CHARA/CLIMB   & E2-S1-W2  & 4 & \object{HD 27777} ($0.204\pm0.006$), \object{HD 31592}\tablefootmark{*}, \object{HD 34053}\tablefootmark{*}\\
            2012-10-20 &  CHARA/CLIMB   & S1-W1-W2  & 3 & \object{HD 31592}\tablefootmark{*}, \object{HD 34053}\tablefootmark{*}\\
            2012-11-27 &  CHARA/CLIMB   & S1-E1-W1  & 5 & \object{HD 32480} ($0.236\pm0.006$), \object{HD 31706} ($0.219\pm0.005$)\\
            2012-11-28 &  CHARA/CLIMB   & S1-E1-E1  & 4 & \object{HD 32480}, ($0.236\pm0.006$) \object{HD 31706} ($0.219\pm0.005$), \\
             & & & & \object{HD 33252} ($0.294\pm0.007$)\\
            2014-11-25 &  CHARA/CLIMB   & E2-S2-W2  & 3 & \object{HD 33252}  ($0.294\pm0.007$)\\  
            2014-11-26 &  CHARA/CLIMB   & E2-S2-W2  & 3 & \object{HD 33252} ($0.294\pm0.007$)\\  
            \hline
            \noalign{\smallskip}
            2009-10-31 & CHARA/CLASSIC & S1-E1 & 4 & \object{HD 32480} ($0.236\pm0.006$)\\
            2009-11-01 & CHARA/CLASSIC & S1-E1  & 3 & \object{HD 32480} ($0.236\pm0.006$), \object{HD 24365}  ($0.319\pm0.008$)\\
            \hline
            \noalign{\smallskip}
            2011-11-07 &  KI   & FT-SEC  & 5 & \object{HD 27777} ($0.204\pm0.006$) \\ \hline
            \noalign{\smallskip} 
            1999-10-09 &  PTI   & NS  & 2 & \object{HD 30111} ($0.555\pm0.056$)\\
            1999-11-03 &  PTI   & NS  & 2 & \object{HD 28024} ($0.222\pm0.026$), \object{HD 27946} ($0.398\pm0.034$), \\
            &&&&\object{HD 25867} ($0.280\pm0.007$)\\
            1999-11-04 &  PTI   & NS  & 1 & \object{HD 28024} ($0.222\pm0.026$), \object{HD 32301} ($0.506\pm0.054$), \\
            &&&&\object{HD 25867} ($0.280\pm0.007$)\\
            1999-12-07 &  PTI   & NS  & 6 & \object{HD 27946} ($0.398\pm0.034$), \object{HD 30111} ($0.555\pm0.056$)\\
            2000-10-16 &  PTI   & NS  & 4 & \object{HD 30111} ($0.555\pm0.056$)\\
            2000-11-13 &  PTI   & NW  & 17 & \object{HD 30111} ($0.555\pm0.056$)\\
            2000-11-14 &  PTI   & NW  & 9 & HD \object{30111} ($0.555\pm0.056$)\\
            2003-10-16 &  PTI   & SW  & 6 & \object{HD 28024} ($0.222\pm0.026$), \object{HD 30111} ($0.555\pm0.056$)\\
            2003-10-22 &  PTI   & SW  & 4 & \object{HD 28024} ($0.222\pm0.026$), \object{HD 30111} ($0.555\pm0.056$), \\
            &&&&\object{HD 27946} ($0.398\pm0.034$), \object{HD 25867} ($0.508\pm0.046$), \\
            &&&&\object{HD 29645} ($0.507\pm0.035$)\\
            2004-10-01 &  PTI   & NW  & 2 & \object{HD 29645} ($0.507\pm0.035$)\\
            2004-10-05 &  PTI   & SW  & 1 & \object{HD 29645} ($0.507\pm0.035$)\\
            [1ex] 
            \hline
        \end{tabular}
        \tablefoot{
        Baselines involving the S2 station (CLIMB/CHARA) on 2012-12-03 produced no (or very faint) interference fringes and so were not suitable for data reduction. All uniform disk (UD) diameters quoted obtained from \citet{Bourges14}.
        \tablefoottext{*}{Calibrator found to be a binary, solution used shown in Section~\ref{AppA}.}
        }
    \end{table*}
    
     \begin{figure*}
        \centering
        \includegraphics[scale=0.6]{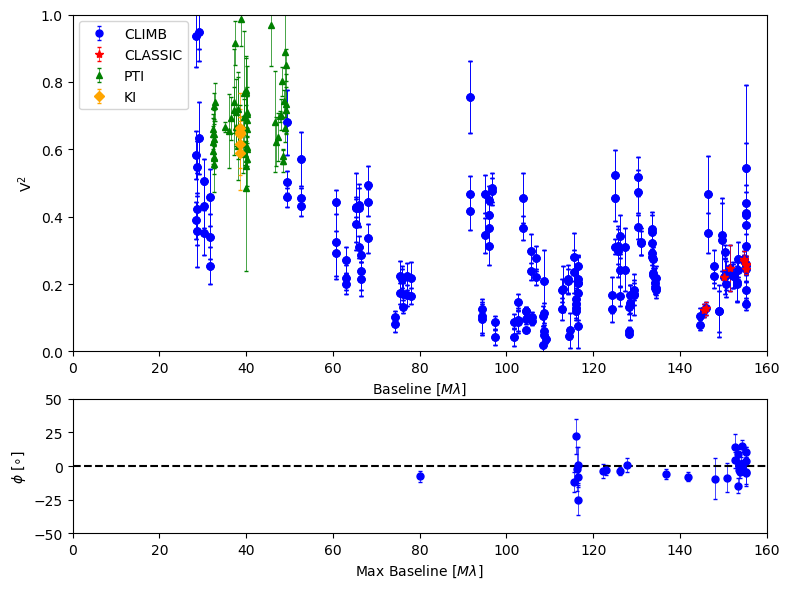}
        \caption{Squared visibility and closure phase measurements against the de-projected baseline length of our interferometric data. Blue data points are those from the CLIMB/CHARA instrument. Red data points are from the CLASSIC/CHARA instrument, Green data is that obtained from the PTI instrument and orange is that from the KI instrument.}
        \label{fig:CPVIS}
    \end{figure*}
    
    Our observations are described in section \ref{lab:observations},  with image reconstruction detailed in section \ref{ImRec}. Our geometrical model fitting approach is described in section \ref{RAPIDO} and radiative transfer modelling and SED fitting are discussed in section \ref{MCMC}. A discussion and analysis of the results can be found in section \ref{Diss}, followed by concluding remarks made in section \ref{Conc}.

\section{Observations} \label{lab:observations}
    The CHARA array is a Y-shaped interferometric facility that comprises six $1\,$m telescopes. It is located at the Mount Wilson Observatory, California, and offers operational baselines between $34$ and $331\,$m \citep{Brummelaar05}. The CLIMB instrument, a three telescope beam combiner \citep{Brummelaar13}, was used to obtain observations in the near-infrared K-band ($\lambda=2.13\,\mu m, \Delta\lambda=0.35\,\mu m$) between October 2010 and November 2014. We obtained 28 independent measurements of SU\,Aur, using seven different 3-telescope configurations with maximum physical baseline of $331\,$m corresponding to a resolution of $\lambda/(2B) = 0.70\,\mathrm{mas}$ [milliarcseconds], where $\lambda$ is the observing wavelength and $B$ is the projected baseline. In addition, a small number of observations were taken in 2009 using the two-telescope CLASSIC beam combiner \citep{Brummelaar13}, also at CHARA in the K-band along the longest ($331\,$m) projected baseline. Details of our observations, and the calibrator(s) observed for the target during each observing session, are summarised in Table~\ref{table:LOG}. The uv\,plane coverage that we achieved for the target is displayed in Figure~\ref{fig:uvplane}. Our data covers a relatively wide range of baseline lengths and position angles, making the data set suitable for image reconstruction.
    
    The CLIMB and CLASSIC data was reduced using pipelines developed at the University of Michigan \citep{Davies18}. This is much better suited to recovering faint fringes from low visibility data than the standard CHARA reduction pipeline of \citep{Brummelaar12}. The measured visibilities and closure phases were calibrated using interferometric calibrator stars observed alongside the target. Their adopted uniform diameters (UDs) were obtained from JMMC SearchCal \citep{Bonneau06, Bonneau11}, when available, or gcWeb\footnote{http://nexsciweb.ipac.caltech.edu/gcWeb/gcWeb.jsp} and are listed in Table\,\ref{table:LOG}. 
    
     \begin{figure*}[t!]
        \centering
        \includegraphics[scale=0.6]{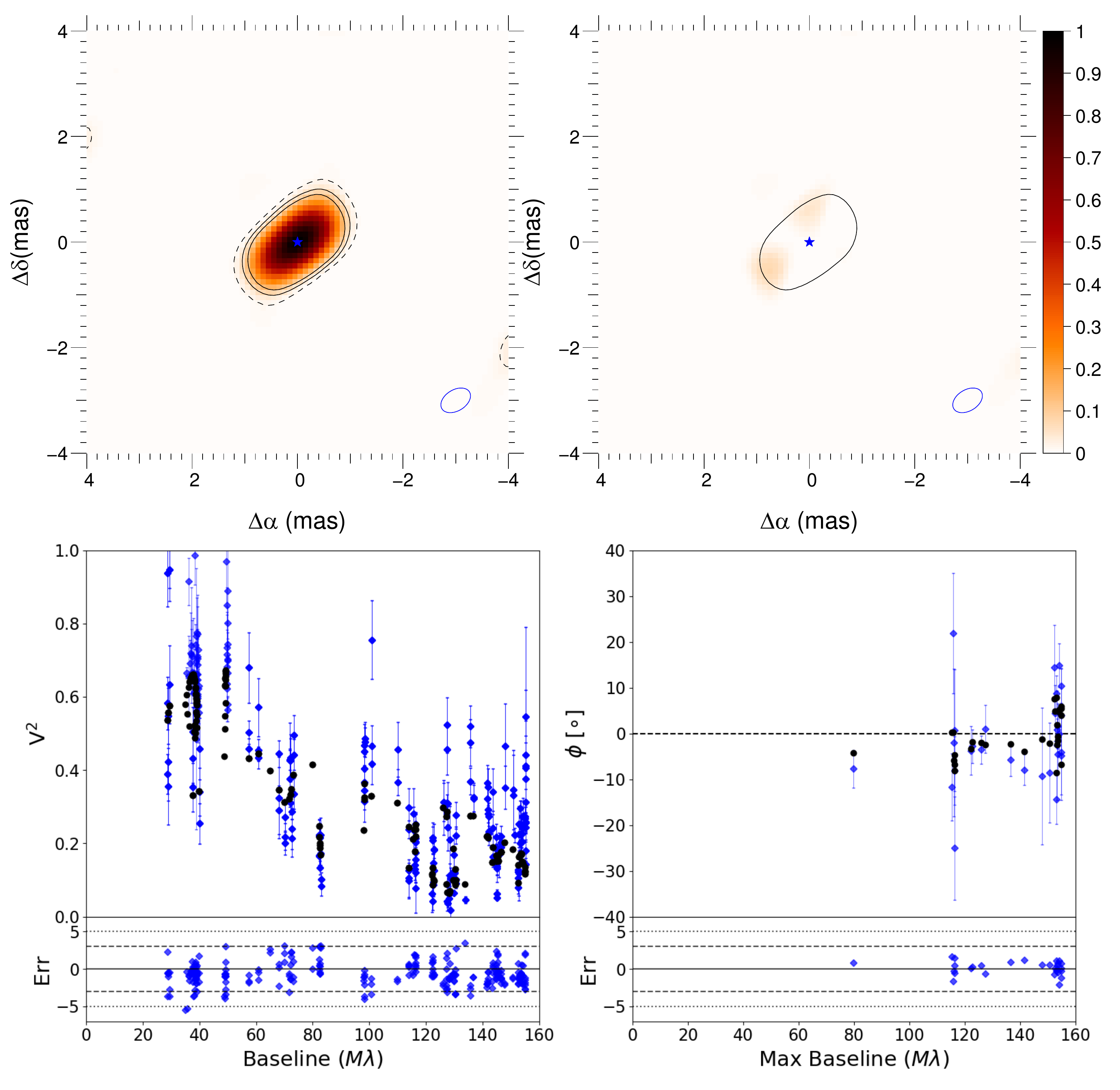}
        \caption{Top left:Resultant reconstructed image showing the $3$ and $5\,\mathrm{\sigma}$ significance levels as solid black lines and the $1\,\mathrm{\sigma}$ level as dashed lines. The beam size is shown in the bottom right. The colour bar is the same for both maps, with the maximum intensity normalised to $1$ for the sake of readability. Top right: Extracted asymmetry in the intensity map shown with $3\,\mathrm{\sigma}$ significance level. Bottom left: Fitted visibilities, data shown in black with model visibility in blue. Bottom right: Fitted closure phases, data shown in black with model CP in blue. The residuals normalised to the standard deviation are plotted in the bottom of each graph.}
        \label{fig:newRec}
    \end{figure*}
    
    During the data reduction it was found that our calibrators HD\,31592 and HD\,34052 exhibited strong closure phase signals, indicating the presence of close companions around these stars. In order to ensure these binary calibrators could be used to calibrate the primary science target, we cross-calibrated the data on these stars with other calibrators observed during the same nights to determine the binary parameters, as outlined in Section~\ref{AppA} in the appendix. Based on the fitted binary parameters, we could then correct the transfer function and use the data for the calibration of SU\,Aur.

    In addition to the new observations from CHARA, archival interferometric data from other facilities was included for our analysis. A small amount of data was available from the Keck Interferometer \citep[KI,][]{Colavita13,Eisner14} from 2011 along a single $84$\,m baseline, while a larger amount of data was also available from the Palomar Testbed Interferometer \citep[PTI,][]{Colavita99} from 1999 to 2004 using a 2 telescope beam combiner on 3 different physical baselines between $86$ and $110$\,m. This data was published in \citet{Akeson05}. These additional measurements complement the CHARA observations in the intermediate baseline range; the full uv coverage is shown in Figure\,\ref{fig:uvplane}.
    
    Both the PTI and KI data were calibrated using the standard method outlined by \citet{Boden98} using the wbCalib software available from NExcScI\footnote{http://http://nexsci.caltech.edu/software/V2calib/wbCalib/}. The calibration pipeline works in conjunction with getCal and the Hipparcos catalogue \citep{Perryman97} for calibrator star astrometry and diameters. This was the same process used by \citet{Akeson05} to extract visibilities from the PTI data. Our re-reduction of this data agrees with the results shown in the literature. The fully reduced data from all instruments is shown in Figure\,\ref{fig:CPVIS}.
    
    As we combine several years worth of data, care was taken to check for time dependencies in the visibilities of baselines of similar length and position angle. Variability in the K band is known to be minimal, so any time dependencies in the visibility amplitudes is likely geometric. However, no significant time dependencies were discovered.

    \begin{table*}[t!]
        \caption{\label{table:geofit}Best fit parameters for the simple geometric models investigated.}
        \centering
        \begin{tabular}{c c c c c} 
            \hline
            \noalign{\smallskip}
             Parameter & Explored Parameter Space & Gaussian &   Ring & Skewed Ring   \\ [0.5ex]
            \hline
            \noalign{\smallskip}
            $R_\mathrm{{min}}$ [mas] & $0.0 - 10.0$ & ... & $1.0\pm0.13$ & $1.1\pm0.12$   \\
            FWHM & $0.0 - 15.0$ & $2.01\pm0.02$ & ... & ...    \\
            $w$ & $0.01-2.0$ & ... & ... & $0.96\pm0.02$  \\
            $\theta$ [$^\circ$] (PA) & $0.0 - 180.0$ & $60.8\pm1.15$  & $61.36\pm1.2$ & $61.0\pm1.0$ \\
            $i$ [$^\circ$]& $0.0 - 90.0$ & $51.4\pm1.04$  & $50.91\pm0.88$ & $51.2\pm1.1$ \\ 
            \hline
            \noalign{\smallskip}
            $\mathrm{\chi^{2}_{red,Vis}}$ & ... & $10.97$  & $11.86$ & $8.57$  \\ 
            \noalign{\smallskip}
            $\mathrm{\chi^{2}_{red,CP}}$ & ... & $1.90$ & $1.90$ & $1.66$\tablefootmark{*} \\ [1ex]
            \hline
        \end{tabular}
        \tablefoot{
        \tablefoottext{*}{The closure phase quoted is the achieved when allowing the skewed ring to become asymmetric. While the software did detect an asymmetry, it failed to constrain its location in the disk. $\theta$ is the minor-axis position angle of the disk.}
        }
    \end{table*}

\section{Image Reconstruction} \label{ImRec}

    Image reconstruction techniques require broad and circular uv coverage along as many baseline lengths as possible. Fortunately, the data from the observations lends itself to this process as the uv plane has been well sampled, though some small gaps remain in the position angle coverage. By comparing visibilities from different instrument at similar baseline lengths and position angles, we can see there is likely very little extended emission in this system and so no correction for instrument field of view is required. This technique is useful for interpretation of non-zero closure phases, indicative of asymmetric distributions, in a model-independent way. Our closure phase values are shown in Figure\,\ref{fig:CPVIS}. There are many different algorithms with which to reconstruct images from interferometric data, but the process described here involved the use of the Polychromatic Image Reconstruction Pipeline (PIRP) which encompasses the $MiRA$ reconstruction algorithm by \citet{MiRA08}. The reconstruction procedure and results are described below.

    In the $MiRA$ routine, the object is modelled as an unresolved central star with an extended, model-independent, environment \citep{Kluska14}. Both components have different spectral behaviours and so differing spectral indices. Additionally, the type and weight of the regularisation was explored, $MiRA$ allows for either quadratic smoothing or total variation regularisations to be implemented. The regularisation plays the role of the missing information by promoting a certain type of morphology in the image. The quadratic smoothing algorithm aims for the smallest possible changes between pixels to produce a smoother image, it is particularly useful as it's quadratic nature means it is less likely to find local minima. On the other hand, total variation aims to minimise the total flux gradient of the image and is useful to describe uniform areas with steep but localised changes. These regularisations are considered to be the best ones for optical interferometric image reconstruction; \citep{Renard11}. 
    The size and number of pixels also plays an important role in image reconstruction. One cannot simply use the maximum number of pixels of the smallest size to obtain better resolution, they have to be chosen to match uv plane sampling. It was found that a quadratic smoothing regularisation with a weight of $1\times10^9$ and $526\times526$ pixels of $0.1$ mas in size provides the best-fit image reconstruction when utilising exact Fourier transform methods. The optimal regularisation parameters were determined using the L-curve method. PIRP also allows for bootstrap iterations \citep{Efron94} starting from a previous best image. This involves a random draw of data points within the dataset to determine the reliable features of the image. The process was carried out 500 times allowing for a pixel by pixel error estimation, see \citet{Kluska16}.
    
    The final image is shown in Figure\,\ref{fig:newRec} (upper left panel), which also shows the asymmetry in the intensity map (upper right panel). The contours represent the $1\sigma$ (dashed line), $3\sigma$ and $5\sigma$ (solid lines) uncertainties. The asymmetry is calculated by rotating the image through $180^\circ$ and subtracting it from the un-rotated image. It is this residual flux that produces the non-zero closure phases, highlighting any areas of greater emission within the disk. Using this technique we can see that the disk has greater intensity in the Eastern regions, which contains $8\%$ more flux that the Western regions with a flux ratio of $1.07$, where the remaining flux is in the central star. By inclining a flared disk with the Western region (bottom right of image) towards the observer the nearside of the inner rim becomes self-shadowed by the near-side disk rim, so the Eastern region inclined away from the observer appears brighter. The best-fit image shows a disk radius of $1.0$~mas, a minor-axis position angle of $41^\circ\pm3$ and an inclination of $51^\circ\pm5$. The fit of the image to the visibility and closure phases in the data is shown in Figure~\ref{fig:newRec}, with a combined visibility and closure phase reduced $\mathrm{\chi^2_{red}}$ of $4.57$.
    
\section{Geometric Model Fitting} \label{RAPIDO}
    The next step in interpreting our interferometric observations is the fitting of simple geometric models to the observed quantities. The visibility profile (Figure~\ref{fig:CPVIS}) reveals a clear drop in visibility through short and intermediate baselines with a possible plateau/second lobe at the longest baselines. 
    In the visibility profile we do not see any evidence for structures on distinct different spatial scales that might indicate the presence of a binary companion or extended halo emission.
    The closure phases are all $<20^\circ$, indicating weak asymmetric features as evidenced in the image reconstruction above. To explore the viewing geometry of the disk, a series of simple intensity distributions were tested against our data.
    All the models tested contained an unresolved point source that was used to represent the central star, a reasonable assumption given the expected angular diameter of the star (see Table.\,\ref{table:Stellar}). The disk component was modelled as one of four intensity distributions: (i) A Gaussian to simulate a disk with unresolved inner rim with a FWHM free parameter. (ii) A ring to simulate emission from a bright inner rim only with a defined fractional width equal to $20\%$ of the radius with an inner radius ($R_{min}$) free parameter. (iii) A skewed ring model with a diffuse radial profile defined by an inner radius ($R_{min}$), a FWHM ($w$) and an azimuthal brightness modulation, where $c_j$ and $s_j$ are the cosine and sine amplitudes for mode $j$, in an attempt to model disk asymmetries. \citet{Lazareff17} find that the diffuse skewed ring here can be used to successfully model a wide range of YSOs. 
    
    With the exception of the skewed ring, these models are intrinsically axisymmetric, but we project the brightness distribution in order to mimic viewing geometry effects that are parameterised with an inclination angle $i$ (defined with $0^\circ$ as face-on) and a disk position angle $\theta$. We measure disk position angles along the minor axis and follow the convention that position angles are measured from North ($\theta = 0^\circ$) towards East. 
   
    The stellar-to-total flux ratio can be calculated by comparing K-band photometry \citep{Curi03} with the stellar atmosphere models of \citet{Kurucz04}, which gives a ratio of $1.17$ (using the stellar parameters listed in Table~\ref{table:Stellar}). However, it cannot be fixed as this has been shown to introduce unreasonable large scale components when fitting long baseline data \citep{Ajay13}. As such the parameter space of the flux ratio was explored step-wise for all models with a range of model parameters. In this way the stellar-to-total flux ratio was constrained for all models. It was found that a ratio of $1.13\pm0.01$ provided the best fit to the data for the ring, skewed ring and TGM models. After determining the star-to-disk flux ratio, the parameter space of the geometric models could be fitted to the observed visibilities and closure phases for each of the observed baselines given initial parameter constraints based on the literature values of \citet{Akeson05} described in Section\,\ref{sec:intro}. We used a bootstrap method to explore the parameter space around these initial values and to compute uncertainties on the individual parameters by fitting Gaussian distributions to parameter histograms.
    
    The best-fit parameters and associated errors for each of the geometric models are shown in Table~\ref{table:geofit}. All test models agree with respect to the position angle and inclination of the disk very well (Table\,\ref{table:geofit}). However, none of the models provide good fits to the data as evidence by the $\chi^2_{red}$ values of between 11.86 (ring model) and $8.57$ (skewed ring). The skewed ring model found a minor axis position angle of $61.0^\circ\pm1.0$ and inclination of $51.2^\circ\pm1.1$. Of the individual models, the skewed ring provides the best fit to both the visibilities and closure phases. Both the Gaussian model and the Skewed Ring are found to be quite diffuse with a FWHM of $2.01~\mathrm{mas}\pm0.02$ and width of $0.96~\mathrm{mas}\pm0.02$ respectively. 

    Overall, we achieved the best-fit with a skewed ring with an inner radius of $1.10~\mathrm{mas}\pm0.12$ and a ring FWHM of $0.96~\mathrm{mas}\pm0.02$, thus making the ring very diffuse with only a marginally defined inner radius. There was no evidence discovered for any over-resolved extended emission or 'halo' found in many other objects \citep{Monnier06a,Kraus2009B}.
    
\begin{figure*}[t!]
        \centering
        \includegraphics[scale=0.915]{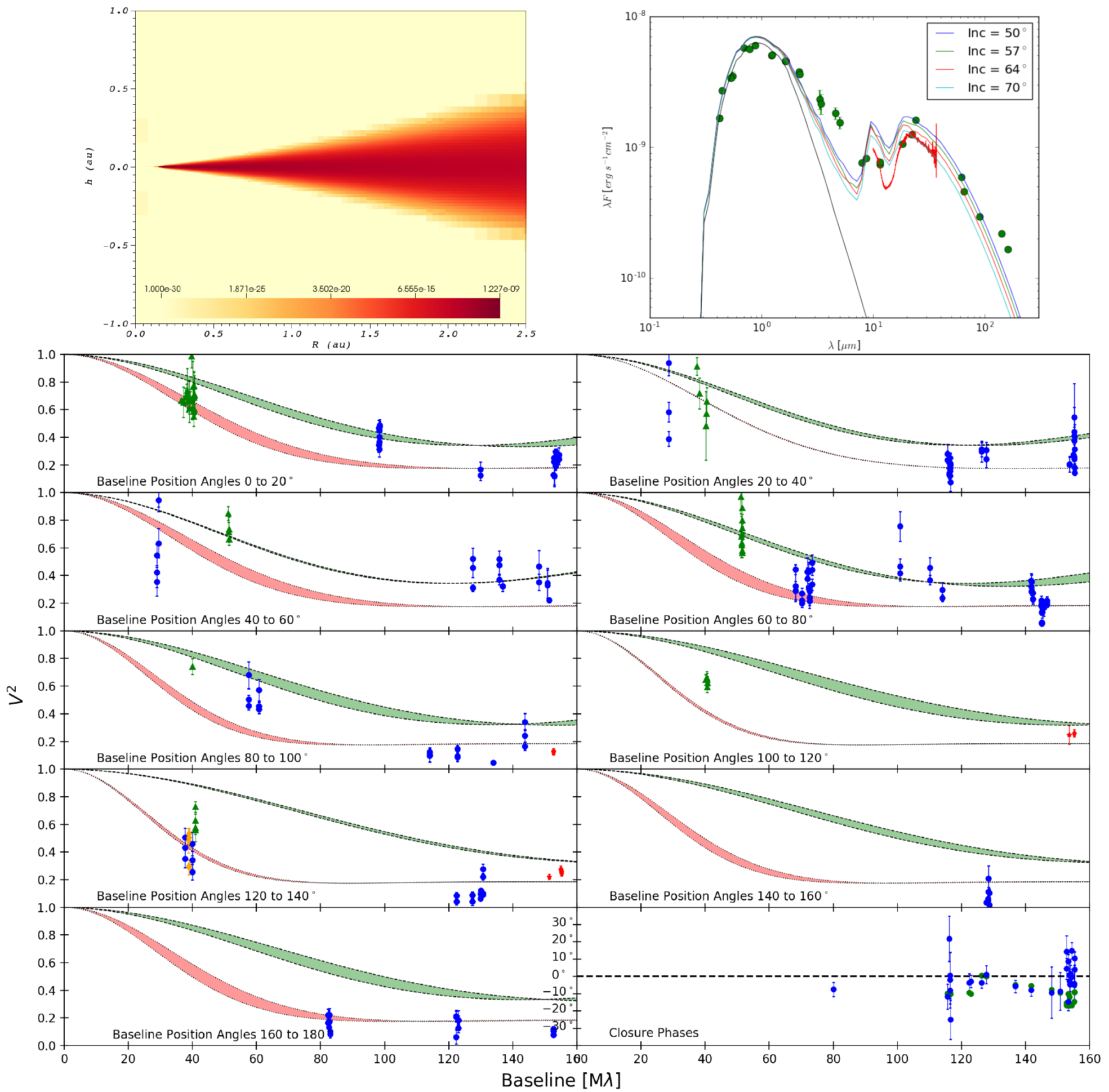}
        \caption{Results of radiative transfer and geometric modelling with the IN05 disk rim prescription \citep{Isella05} of a gas density-dependent sublimation temperature with a single grain species size $0.1 \mathrm{\mu m}$. TOP LEFT: Disk density cross section of the inner rim. A logarithmic colour scale is used with a minimum density of $1.00\times10^{-30}$ to a maximum of $4.64\times10^{-10}$. TOP RIGHT: SED computed with our radiative transfer model. Dark blue curve is the simulated blackbody emission of the central star. Green points are photometric observations while the short red line is the \textit{Spitzer} spectrum. The coloured curves represent the SED at the different inclinations of 50, 57, 64 and $70^\circ$. BOTTOM: Visibility data binned by position angle of observation. The data points are split by instrument consistently with Figures\,\ref{fig:uvplane} and\, \ref{fig:CPVIS}, where blue circles are from CHARA/CLIMB, red stars are from CHARA/CLASSIC, green triangles from PTI and orange diamonds from KI. The red curves are the results of the best fit geometric skewed ring model and the green curves are calculated from the radiative transfer image at an inclination of $50^\circ$ and a position angle of $60^\circ$. The dashed bounding lines indicate the minimum and maximum model visibilities for that position angle bin. The very bottom right panel shows the observed CP measurements (black) and the CP computed from the radiative transfer image (green).}
        \label{fig:IN05}
    \end{figure*}

\begin{figure*}[t!]
        \centering
        \includegraphics[scale=0.915]{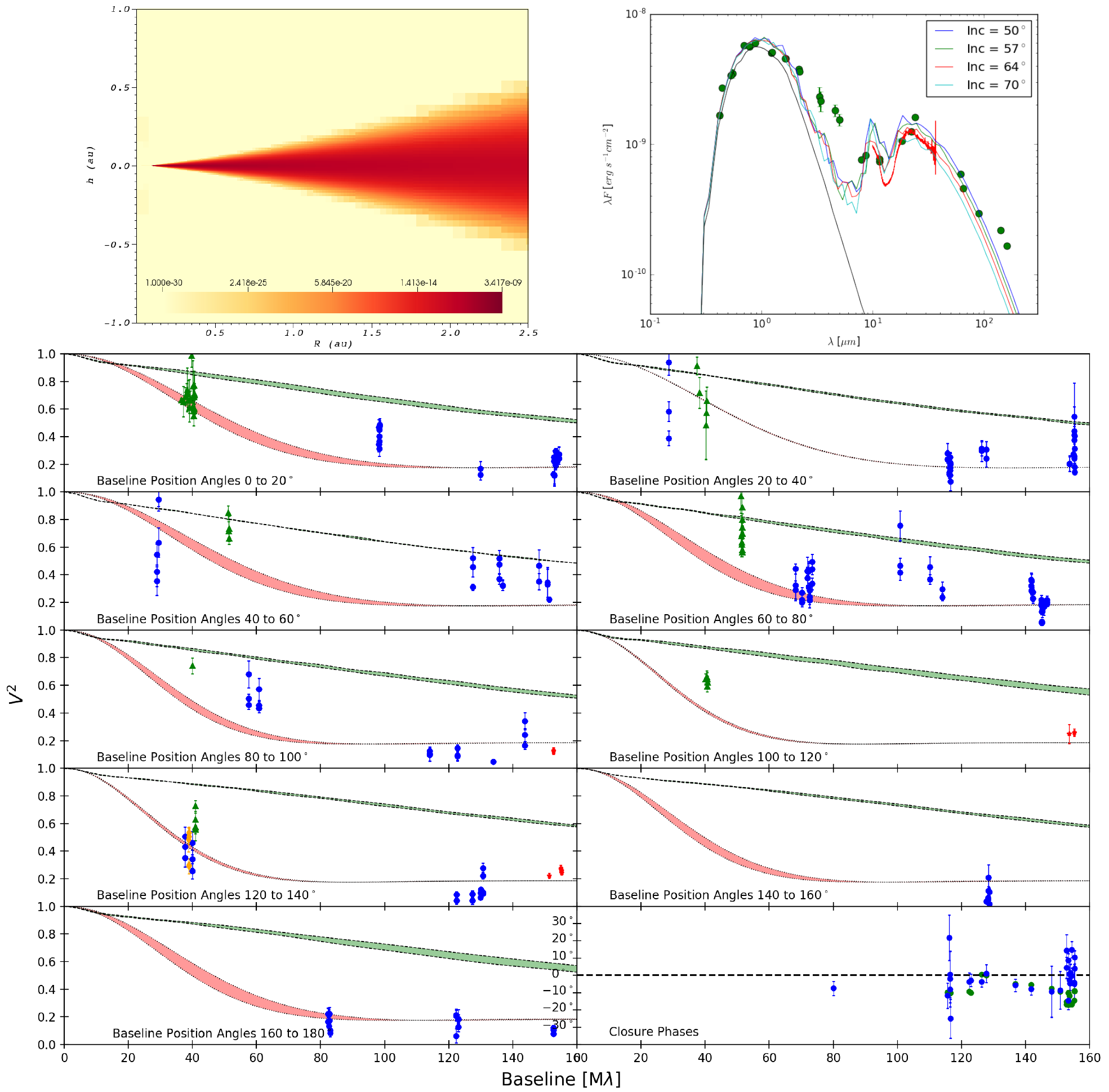}
        \caption{Results of radiative transfer and geometric modelling with the THM07 disk rim prescription \citep{Tannirkulam07} of a gas density-dependent sublimation temperature with two grain species, a majority larger grains at $1.2 \mathrm{\mu m}$ with fewer smaller grains at $0.1 \mathrm{\mu m}$. The mass of larger grains is fixed at $9$ times the mass of smaller grains. TOP LEFT: Disk density cross section of the inner rim. A logarithmic colour scale is used with a minimum density of $1.00\times10^{-30}$ to a maximum of $1.23\times10^{-09}$. TOP RIGHT: SED computed with our radiative transfer model. Dark blue curve is the simulated blackbody emission of the central star. Green points are photometric observations while the short red line is the Spitzer spectrum data. The coloured curves represent the SED at the different inclinations of $50$, $57$, $64$ and $70^\circ$. BOTTOM: Visibility data binned by position angle of observation. The data points are split by instrument consistently with Figures\,\ref{fig:uvplane} and\, \ref{fig:CPVIS}, where blue circles are from CHARA/CLIMB, red stars are from CHARA/CLASSIC, green triangles from PTI and orange diamonds from KI. The red curves are the results of the best fit geometric skewed ring model and the green curves are calculated from the radiative transfer image at an inclination of $50^\circ$ and a position angle of $60^\circ$. The dashed bounding lines indicate the minimum and maximum model visibilities for that position angle bin. The very bottom right panel shows the observed CP measurements (black) and the CP computed from the radiative transfer image (green). }
        \label{fig:THM07}
    \end{figure*}

\begin{figure*}[t!]
        \centering
        \includegraphics[scale=0.915]{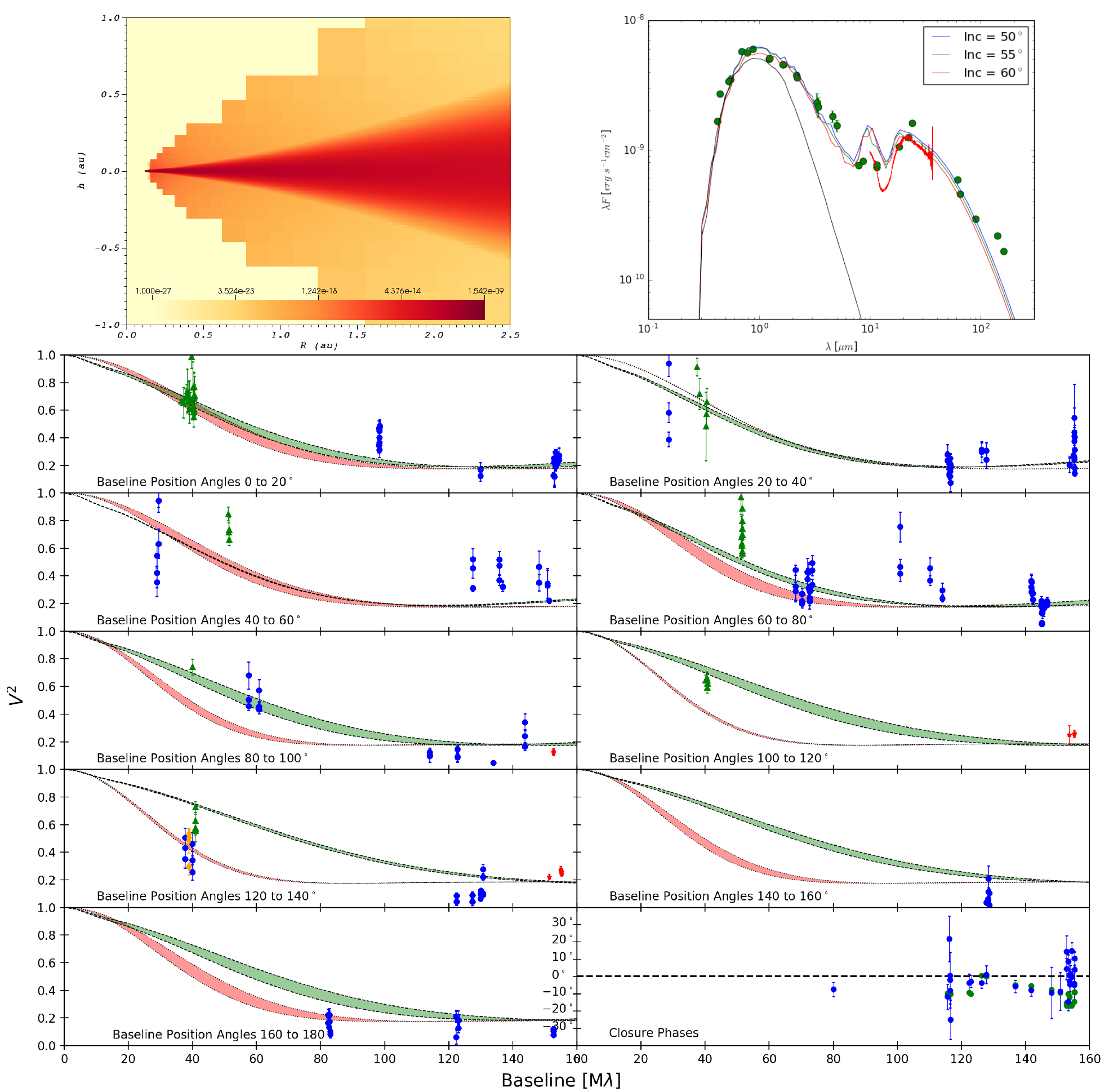}
        \caption{Results of radiative transfer and geometric modelling with the IN05 disk rim prescription \citep{Isella05} and an additional dusty disk wind (BK12) implemented following \citet{Bans12}. TOP LEFT: Disk density cross section of the inner rim. A logarithmic colour scale is used with a minimum density of $1.00\times10^{-27}$ to a maximum of $1.54\times10^{-09}$. TOP RIGHT: SED output of radiative transfer analysis. Dark blue curve is the simulated blackbody emission of the central star. Green points are photometric observations while the short red line is the Spitzer spectrum data. The coloured curves represent the SED at the different inclinations of $50$, $55$ and $60^\circ$. BOTTOM: Visibility data binned by position angle of observation. The data points are split by instrument consistently with Figures\,\ref{fig:uvplane} and\, \ref{fig:CPVIS}, where blue circles are from CHARA/CLIMB, red stars are from CHARA/CLASSIC, green triangles from PTI and orange diamonds from KI. The red curves are the results of the best fit geometric skewed ring model and the green curves are calculated from the radiative transfer image at an inclination of $50^\circ$ and a position angle of $60^\circ$. The dashed bounding lines indicate the minimum and maximum model visibilities for that position angle bin. The very bottom right panel shows the observed CP measurements (black) and the CP computed from the radiative transfer image (green).}
        \label{fig:BK12}
    \end{figure*}

    In order to model the observed CP signal, with a maximum of $20\pm12^\circ$, which may indicate a slight asymmetry in the disk, we introduced asymmetries to the skewed ring model \citep{Lazareff17}. However, this improved the fit only marginally over the standard ring model from Table~\ref{table:geofit}. It was found that zero closure phases produced a reduced chi-squared ($\mathrm{\chi^2_{red,CP}}$) fit of $1.90$, while a sinusoidally modulated asymmetric ring only resulted in a $\mathrm{\chi^2_{red,CP}}$ of $1.66$. The model visibility curves corresponding to the best-fit skewed ring model are shown in Figure~\ref{fig:IN05}, \ref{fig:THM07} and \ref{fig:BK12} (red curve). The different panels show visibilities towards different position angle bins. 
    
    Of the simple geometric models tested, the skewed ring model provides the best fit. However, non can be said to provide a good fit to the observed data, as evidence by the $\chi^2_{red}$ values shown in Table\,\ref{table:geofit}. As such, more complex disk structures are required, such as flared disks, different rim morphologies or disk winds. In the next section, these possibilities are explored in detail using radiative transfer techniques. This allows us to not only explore complex geometries, but to derive physical parameters such as radial density profiles and the disk scale heights. 
    
\section{Radiative Transfer Modelling}\label{MCMC}

    We used the TORUS Monte-Carlo radiative transfer code \citep{Harries00}, allowing for the simultaneous fitting of visibility, closure phase and photometric data to further constrain the geometry and physical dust properties of the SU\,Aurigae circumstellar disk.

    Starting from the disk properties derived by  \citet[][Table~\ref{table:torus}]{Akeson05}, we explored radiative transfer models with different scale heights (where scale height is that of the gas parameterised at $100~\mathrm{au}$ with flaring index $\beta$) and inner rim shapes. In our TORUS simulations, the dust was allowed to vertically settle and the dust sublimation radius was left as a free parameter, allowing the inner rim radius to define itself based on well-defined rules of the \citet{Lucy99} iterative method to determine the location and the temperature structure of the whole disk. This is implemented whereby the temperature is initially calculated for grid cells in an optically thin disk structure, with dust added iteratively to each cell with a temperature lower than that of sublimation, until the appropriate dust to gas ratio is reached ($0.01$). Once TORUS has converged to radiative equilibrium a separate Monte Carlo algorithm is used to compute images and SEDs based on the optical properties of the dust species implemented. We confirmed that stellar photosphere models of \citet{Kurucz04} using these stellar parameters can reproduce the photometry measurements of SU\,Aur reasonably well across the visible continuum. The grain size distribution used by \citet{Akeson05} is a distribution of astronomical silicate grains up to $1\,\mathrm{mm}$ in size. We adopt a silicate grain species with dust properties and opacities adopted from \citet{Draine84}. The initial density structure of the gas is based upon the $\alpha$-disk prescription of \citet{Shakura73} where the disk density is given as:
    
    \begin{equation}\label{eq:1}
    \centering
     \rho(r,z) = \frac{\Sigma(r)}{h(r)\sqrt{2\pi}}\exp\bigg\{-\frac{1}{2}\bigg[\frac{z}{h(r)}\bigg]^2\bigg\} .
    \end{equation}
    
    Here, $z$ is the vertical distance from the midplane while the parameters $h(r)$ and $\Sigma(r)$ describe the scale height and the surface density respectively. 
    
    In our radiative transfer models, we represent the stellar photosphere with a \citet{Kurucz79} model atmosphere using the stellar parameters outlined in Table~\ref{table:Stellar}.
    The photometric data was obtained from a wide range of instruments from the VizieR database and are compiled in Table~\ref{table:Photometry}. Where multiple observations in the same waveband were present care was taken to minimise the total number of instruments and keep the number of observation epochs as close as possible to minimise any potential variability effects. 

\begin{figure}[b]
        \centering
        \includegraphics[scale=0.95]{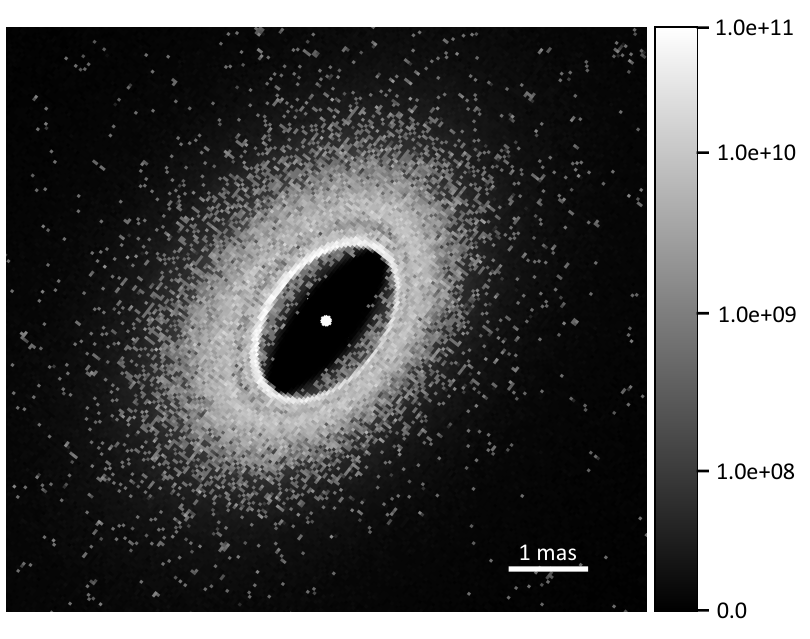}
        \caption{Computed synthetic image from TORUS following the BK12 dusty disk wind prescription. The colour indicates intensity in the K-band.}
        \label{fig:Timage}
    \end{figure}

    Visibilities were calculated from synthetic images of the disk system (as shown in Figure~\ref{fig:Timage}) extracted through application of the van Cittert-Zernicke theorem, applied using a 1D Fourier transforms projected onto the observed baseline position angle. Phases are also extracted from the images and are used to calculate the closure phase, as described in \citet{Davies18}.
    
    The parameter space of the radiative transfer models was explored objectively using the values of \citet{Akeson05} as a starting point. A range of physically realistic values for each parameter was explored in a broad grid of models (as described in Table~\ref{table:torus}). A $\chi^2$ value was then computed for the visibilties, closure phases and SED fits of each model allowing the grid to be refined around the minimum. The interferometric and photometric data points were fitted simultaneously, with the resulting $\chi^2$ values shown in Table~\ref{table:Chi2}. The silicate feature at $10\,\mathrm{\mu m}$ allows us to place some constraints on the dust sizes, as larger grains produce smaller features. The growth of dust grains and their effects on observed silicate features is described in a review by \citet{Natta07}. The IR flux is controlled by the morphology of the sublimation rim. As the inner radius increases, the amount of circumstellar material emitting in mid-IR wavelengths is reduced, and the IR emission decreases. The shape of the mid-IR excess also describes the degree of flaring present in the disk, where large excess indicates greater flaring. A larger flaring power in a disk leads to an increasing surface intercepting the starlight, and therefore an increase in reprocessed radiation. 
    We adjusted the total dust mass in the model in order to match the millimeter flux. A detailed description of the effect of disk parameters on the SEDs of protoplanetary disks can be found in \citet{Robitaille07}.
    
    Due to the optical depth of the system the inner-rim of the disk appears as the brightest part of the disk at NIR wavelengths. This is because the rest of the disk is shadowed by the inner rim and only rises out of shadow in cooler regions of longer wavelength emission.

    \subsection{Sublimation Rim Model} \label{SubRim}
 
    The curved rim of \citet{Isella05}, henceforth IN05, is based upon a single grain size prescription with a gas density-dependent sublimation temperature. Due to a vertical gas pressure gradient, the sublimation temperature decreases away from the mid-plane, creating a curved rim. The sublimation temperature of the grains follows
    \begin{equation}
        T_{sub} = G\rho^\gamma(r,z),
    \end{equation}
    where the constant $G=2000$~K, $\gamma = 1.95\times 10^{-2}$, $r$ is the radial distance into the disk and $z$ is the height above the midplane \citep{Pollack94}. 
    
    A curved rim is shown to be a viable disk model by \citet{Flock16a} and \citet{Flock16b}, based upon extensive hydrodynamical simulations. A wide range of disk structure parameters are explored to find the best fit solution to both the interferometric measurements and the SED.
    Figure~\ref{fig:IN05} shows the results of radiative transfer modelling of the IN05 rim prescription. The model SED shows a clear flux deficit in the NIR, with a K-band flux of just 64\% of the 2MASS photometric point, far outside the limited range of variability of SU\,Aur. This is also clear in the visibility curves, shown in green, where the overall shape is a good fit, but the minimum visibility is too high due to a larger than expected stellar contribution.
    
    An alternative sublimation front geometry is proposed by \citet{Tannirkulam07}, henceforth THM07. This model employs a two-grain scenario, were a mixture of small $0.1\,\mathrm{\mu m}$ grains and large $1.2\,\mathrm{\mu m}$ grains has been adopted, with the mass of larger grains fixed at $9$ times the mass of smaller grains. The smaller grains are not allowed to settle, so the scale height is fixed to that of the gas. The larger grains are allowed to settle to 60\% of the scale height of the gas. This combined with the larger grains existing closer to the star, due to more efficient cooling, leads to an elongated and curved sublimation front. Figure~\ref{fig:THM07} shows the results of this modelling. The SED again shows a clear deficit in NIR flux, with a K-band flux of just 68\%, comparable with that of the IN05 prescription. The presence of dust closer to star changes the shape of the visibility curve dramatically, with the first lobe now extending to much longer baselines. The two grain THM07 model proves to be a worse fit than the single grain IN05 model.
    
    Our results show that the best fit of the curved-rim disk model of IN05 can be achieved with a single silicate grain species with a single grain size of $0.1\,\mu m$. The addition of larger grain species further reduced near-infrared flux in 1-3\,$\mu m$ region, resulting in a poorer fit to the SED in both the shape and magnitude of the NIR excess. The disk is also found to be highly flared and extending from $0.15$ to $100$\,au, loosely constrained by the long wavelength photometry, while still solving for vertical hydrostatic equilibrium. Importantly, comparison of the stellar atmosphere, represented by the Kurucz model atmosphere \citep{Kurucz04}, with the photometric data can provide the stellar-to-total flux ratio for each waveband. In the K-band this ratio is found to be $1.27$, meaning the circumstellar environment contributes $44\%$ of the total flux. The dust to gas ratio is fixed to $0.01$, as taken from literature values \citep{Akeson05}. The NIR flux deficit results also in a poor fit to the K-band visibilities.

    \subsection{Dusty Disk Wind Model}
    
    In an attempt to increase the NIR flux contributions in the model, we explored a dusty disk wind scenario as set out by \citet{Bans12}, henceforth BK12. This mechanism is based on the presence of a large-scale, ordered magnetic field which threads the disk. The field could originate from the interstellar field that permeates the molecular cloud and is dragged by in-falling gas into the disk. The magnetic field strength required to drive these winds is of the order of kGauss. Fields of this strength have been shown to be present in other T~Tauri stars ($2.35\pm0.15\,\mathrm{kG}$ for T~Tau) by \citet{Guenther99} and \citet{Krull99}. In this model material is flung out along magnetic field lines highly inclined to the disk surface. The high magnetic pressure gradient above the disk surface accelerates the material which is then collimated through the azimuthal and poloidal field components \citep{Bans12}. These centrifugally driven winds are highly efficient at distributing density above and below the plane of the disk, carrying angular momentum away from the disk surface.

    \begin{table*}[t!]
        \caption{\label{table:torus}Best-fit parameters resulting from SED and visibility fitting.
        }
        \centering
        \begin{tabular}{c c c c c} 
            \hline
            \noalign{\smallskip}
            Parameter  & Literature value & Reference & Range explored &  Best fit value    \\ [0.5ex]
            \hline
            \noalign{\smallskip}
            $i$ & $60^\circ$ & (1),(2) & $30 - 80^\circ$ & $50^\circ$    \\
            $R_\mathrm{{inner}}$ & $0.18$ & (1)(3) & $0.1 - 0.6~\mathrm{au}$ & $0.15~\mathrm{au}$    \\
            $R_\mathrm{{outer}}$ & $100~\mathrm{au}$ & (1) & $20.0-120.0~\mathrm{au}$ & $100.0~\mathrm{au}$    \\
            $h_\mathrm{0}$ & ... & ... & $7.0 - 20.0~\mathrm{au}$ & $15.0~\mathrm{au}$     \\
            $\alpha_\mathrm{{disk}}$ & ... & ... & $1.0 - 3.0$ & $2.4$  \\
            $\beta_\mathrm{{disk}}=(\alpha-1)$ & ... & ... & $0.0 - 2.0$ & $1.4$  \\
            $\mathrm{Dust:Gas}$ & $0.01$ & (1) & $0.01 - 0.008$ & $0.01$    \\
            $a_\mathrm{{min}}$ & $0.1\,\mathrm{\mu m}$ & (1) & $0.1-1.4~\mathrm{\mu m}$ & $0.39~\mathrm{\mu m}$ \\  
            $q_\mathrm{{dist}}$ & $3.0$ & (1) & $2.00 - 4.00$ & $3.06$  \\
            $T_\mathrm{{sub}}$ & $1600$ & (1) & $1400- 2000$ & $1600~\mathrm{K}$ \\ [1ex] 
            \hline
            \noalign{\smallskip}
            Dusty disk wind parameter & Literature value & Reference & Range explored & Best fit value \\
            \hline
            \noalign{\smallskip}
            $R_\mathrm{{0min}}$ & ... & ... &  $2.0 - 10.0~\mathrm{R_{\sun}}$ & $4.5~\mathrm{R_{\sun}}$ \\
            $T_\mathrm{{wind}~(near~surface)}$ & ... & ... & $1400 - 2400~\mathrm{K}$ & $1600~\mathrm{K}$ \\
            $\mathrm{Opening~Angle}$ & ... & ... & $25 - 55^\circ$ & $45^\circ$ \\
            $\dot{M}$ & ... & ... & $10^{-6} - 10^{-12}~\mathrm{M_{\odot}yr^{-1}}$ & $10^{-7}~M_{\odot}yr^{-1}$ \\
            \hline
            
        \end{tabular}
        \tablebib{
        (1) \citet{Akeson05}; (2)~\citet{Unruh04}; (3)~\citet{Jeffers14}
        }
        \tablefoot{
        Computed using the TORUS radiative transfer model \citep{Harries00} for the BK12 model scenario. $q_\mathrm{{dist}}$ is the power law of the grain size distribution. The best fit values are those of the BK12 prescription rather than the IN05 or THM07 models. $\alpha_\mathrm{{disk}\,\,is\,\,fixed\,\,at\,\,(\beta_{disk}+1)}$.
        }
    \end{table*}
    
    The BK12 wind is launched from the disk surface in a region between an inner and outer radius, located at a maximum of twice the sublimation radius. It is FUV radiation that plays a key role (through photoevaporation) in the mass loading of the wind at the inner launching region. The gas density structure is given as
    \begin{equation} \label{eq:2}
    \rho = \rho_1\bigg(\frac{r_0}{r_1}\bigg)^{3/2} \eta(\chi),
    \end{equation}
    where $\rho$ is the density along a flow line at a distance $r_0$ from the symmetry axis, i.e. along this disk surface. $\rho_1$ denotes the density at the fiducial radius of $r_1$, which is given as $1$\,au while $\chi$ is related to the cylindrical coordinate $z$ through
    \begin{equation}\label{eq:3}
       \chi = z/r .
    \end{equation}
    $\eta(\chi)$ is obtained from the solution of the MHD wind equations. The into-wind mass transfer is linked to the density through
    \begin{multline}
     \rho = 1.064\times10^{-15}\Big(\frac{\dot{M}_{out}}{10^{-7}M_\odot yr^{-1}}\Big)\Big(\frac{M_*}{0.5M_\odot}\Big)^{-1/2}  \\ \times \frac{1}{\mathrm{ln}(r_{0max}/r_{0min})\psi_0(1-h_0\xi_0^{'})} \mathrm{~g~cm^{-3}},
    \end{multline}
    where $\dot{M}_{out}$ is the mass outflow rate, $M_*$ is the stellar mass, $h_0$ is the disk scale height and $\psi_0$ is the ratio of vertical speed to Keplarian speed at the disks surface. $\xi_0^{'}$ is related to the angle $(\theta_0)$ at which the poloidal component of the magnetic field threads the disk, defining the opening angle of the disk wind \citep{Safier93}
    \begin{equation}
    \xi_0^{'} = tan(\theta_0).
    \end{equation}
    This rate generally controls the magnitude of the NIR excess added by the dusty wind.
    
    This prescription is taken from \citet{Safier93}, and assumes a steady, axisymmetric, effectively cold disk outflow. For the dust distribution, a constant dust-to-gas ratio is assumed to match that of the disk. The wind is populated with dust in the same way as the disk and also converges towards radiative equilibrium with each \citet{Lucy99} iteration. Full details of the disk wind model implemented can be found in \citet{Bans12} and \citet{Konigl11}. A wide parameter search was undertaken, in order to determine the optimum solution (see Table~\ref{table:torus}). The stellar parameters were fixed to those shown in Table~\ref{table:Stellar} and the same silicate prescription was adopted for all models, as in the IN05 and THM07 prescriptions The position angle of the disk was fixed to that of the best fit geometric model listed in Table~\ref{table:geofit}.
    
    The results of our radiative transfer modelling of the BK12 wind are shown in Figure~\ref{fig:BK12}. The cross section shows the very inner-rim of disk, uplifted dust above and below the mid-plane can clearly be seen. The inner-rim is also curved using the IN05 prescription, although a grain size of $0.4\,\mathrm{\mu m}$ is required to produced the observed excess across the infrared. A range of grain sizes were tested from $0.1$ to $1.5\,\mathrm{\mu m}$ but $0.4\,\mathrm{\mu m}$ grains provided the best fit to the K-band photometric flux values. 
    
    An into-wind mass outflow rate of $1\times10^{-7}\mathrm{ M_{\odot}yr^{-1}}$ was required to uplift enough material to reproduce the observed excess. Lower into-wind outflow rates do not allow enough material to exist exterior to the inner-rim to reproduce the observed K-band excess in the photometry and interferometric data. If one assumes an outflow to accretion ratio of $0.1$ one can predict an accretion rate of $1\times10^{-6}\mathrm{ M_{\odot}yr^{-1}}$. This is unphysically high for an object such as SU\,Aur given the expected age of the system ($5.18\pm0.13~\mathrm{Myr}$), assuming that the rate stayed constant over full period. In addition, this is in disagreement with non-detection of $\mathrm{Br\gamma}$ emission by \citet{Eisner14} which is suggestive of a lower outflow/accretion rate. Other free parameters in this parameterisation of a dusty disk wind cannot reproduce the effect of a high into-wind mass accretion rate.
    
    The dust wind scenario has the effect of increasing the amount of the dust close to the star, where temperatures are sufficient for NIR flux contribution. As shown in Figure~\ref{fig:BK12} this model is in agreement with the observed NIR photometry, with a K-band flux of 102\% of the photometric value, well within the range of variability of SU\,Aur. The rest of the SED is still well fitted, as the curvature and shape are based upon the IN05 prescription described above. The visibility curves shown provide a good fit to the data points, matching the lowest visibilites well. However, one cannot negate the potentially unphysically high outflow/accretion rates required to successfully model the data. The introduction of disk wind also had the interesting effect of flattening the 'bump' in the second visibility lobe present in many disk models, including the IN05 prescription. The $\chi^2$ fits for the visibilities and SED for all three models are shown in Table\,\ref{table:Chi2}.

    \begin{table}[b]
        \caption{\label{table:Chi2} $\chi^2$ results of radiative transfer fits to the squared visibilities, the closure phases (CP) and the SEDs. }
        \centering
        \begin{tabular}{c c c c} 
            \hline
            \noalign{\smallskip}
            Model  &  $\chi^{2}_{V^2}$ & $\chi^{2}_{CP}$ & $\chi^{2}_{SED}$   \\ [0.5ex]
            \hline
            \noalign{\smallskip}
            IN05 & 91.9 & 0.1 & 151.1  \\
            THM07 & 84.2 & 0.1 & 137.7 \\
            BK12 & 35.6 & 0.1 & 121.3 \\ [1ex] 
            \hline
        \end{tabular}
        \tablefoot{
        $\chi^2$ values are only reduced by the number of data points due to complexity of the degrees of freedom in TORUS.
        }
    \end{table}

\section{Discussion}\label{Diss}
    
    In investigating the circumstellar environment of SU Aur we have explored the structure and composition of the disk and greatly improved the constraints on the parameters initially taken from literature. The wide variety of techniques used to analyse the interferometric data allow us to precisely define the disk characteristics.
    
    Image reconstruction shows a disk inclined at $52.8^\circ\pm2.2$, this is in agreement with values of $63^{\circ+4}_{-8}$, \textasciitilde$60^\circ$ and \textasciitilde$50^\circ$ found by \citet{Akeson05,Unruh04,Jeffers14} respectively. The minor axis $\theta$ on the other hand was found to be $50.1^\circ\pm0.2$, greater than the literature values of $24^\circ\pm23$ and $15^\circ\pm5$ found by \citet{Akeson05,Jeffers14}. This difference is likely due to either: The poor uv coverage and lack of longer baselines in previous interferometric studies, both of which make estimating the position angle and inclination particularly unreliable. Other non-interferometric studies focus on the outer disk, rather than the inner au-scale regions.
    The image reconstruction also reveals evidence of slight asymmetries within the disk at $3$ sigma significance level, that are consistent with an inner disk rim seen at an intermediate inclination. As the K-band emission primarily traces the very inner region of the rim; if a disk is inclined the near side of the rim will be partially obscured from view, while the far side of the rim will be exposed to observation. This explanation can successfully account for the over-brightness observed in the Eastern disk region in the image reconstruction shown in Fig.\,\ref{fig:newRec} and can also be seen in the radiative transfer image shown in Fig.\,\ref{fig:newRec}. Models of the effect of inclined disk on the observed brightness distribution are described by \citet{Jang13}. There are several other scenarios that have been used to explain asymmetries in protoplanetary disks in the past. Two possible scenarios are: Firstly, a shell ejection episode that can carry dust and gas away from the central star can be capable of reproducing the photometric variability in different epochs of observations \citep{Fernandes09,Kluska18}. Also, the presence of a companion embedded within the disk can create dust trapping vortices that capture dust grains. These vortices, however, are known to trap primarily large grains (mm-size), not the small micron-sized grain we observe in the infrared, as shown by \citet{Kraus17,vdMarel13}. We rule out the presence of a companion by undertaking a companion search using our geometric models; a second point source was iterated through the parameter space with a grid size of $100$~mas in steps of $0.1$~mas (see Sec.\,\ref{ImRec}). However, the model fit did not improve significantly by adding an off-center point source. We therefore favour the explanation of asymmetry arising from an inclined disk. 
    
    Our geometric model fits were key in understanding the circumstellar environment of SU\,Aur. It was found that a skewed ring structure is able to fit our data best, which suggests that we trace a diffuse inner disk edge. This finding is consistent with the studies on other YSOs that found bright inner rims, such as the \citet{Lazareff17} survey of 51 Herbig AeBe stars using the PIONIER instrument at the VLTI. They found that over half of the disks could be successfully modelled using a diffuse ring structure. A high optical depth of the circumstellar material causes a bright inner rim, where most of the radiation is absorbed, scattered or re-emitted. Our best-fit Skewed ring model suggests that the stellar-to-total flux ratio is $1.13$ and achieves a $\mathrm{\chi^2_{red}}$ of $8.57$. \citet{Akeson05} find that $44\pm9\%$ of the total flux is in the SED K-band excess, with a $4\%$ of flux in an extended envelope, this is in good agreement with the values found from geometric modelling. The skewed ring model fits a ring of radius $0.17\pm0.02$\,au at an inclination of $51.2^\pm1.1^{\circ}$. These values are in remarkable agreement with both the values derived from image reconstruction in this study and the literature values of \citet{Akeson05} of $0.18\pm0.04$\,au and $62^{\circ+4}_{-8}$, where the slight differences are likely due to difference between the diffuse profile, skewed ring and the standard ring structures employed. The minor axis $\theta$ of $61.0\pm1.0^\circ$ is similar to the image reconstruction value, which is significantly larger than literature values. As above, this is most likely due to the poor uv coverage and short baselines available in previous interferometric studies, making the estimates of position angle and inclination particularly unreliable. The skewed ring also introduces modulated asymmetries into the ring profile.
    The skewed ring geometric model with azimuthal brightness modulation results in an improved fit with $\mathrm{\chi^2_{red}}=1.66$, where the contrast of the asymmetry is consistent with the one found using image reconstruction techniques. Similarly this can be attributed to inclination effects of the viewing geometry \citep{Jang13}. However, the position angle of the asymmetry is not well constrained in these models.
    
    Radiative transfer modelling of the disk allowed us to fit a physical disk model to the visibility and photometry data simultaneously, meaning the 3-D density distribution of the disk can be explored. In this paper, three different geometries are considered: The single grain curved rim of IN05, the two grain curved rim of THM07 and the addition of a dusty disk wind of BK12 to the single grain curved rim. In the case of the IN05 prescription, we follow the idea that the sublimation temperature is gas-pressure dependent allowing the rim shape to be defined as described in Sec.\,\ref{SubRim}.  This model provides good constraints on the characteristic size of the near-infrared emitting region and the flaring in the colder regions, with a sublimation temperature of $1600$\,K corresponding to an inner radius of $0.12$\,au, slightly smaller than than the literature values of $0.18\pm0.04$\,au \citep{Akeson05} and $0.17\pm0.08$\,au \citep{Jeffers14}. The disk was also found to be at an inclination of \textasciitilde$50^\circ$ and position angle of \textasciitilde$45^\circ$, in agreement with both the literature and above mentioned methods. This is shown to fit the photometry well at both shorter and longer wavelengths. However, there is a clear deficit in the IR excess, which also leads to poorly fitted visibilities due to an over-estimation of the stellar-to-total flux ratio. The same issue is obvious in the THM07 prescription, where larger grains are introduced allowing dust to exist closer to the star, though this model also fails to reproduce the visibility curve of the geometric modelling, as the inner radius of the disk is much smaller. A two grain model does not provide a good fit to these observations. 
    
    The dusty disk wind prescription of BK12 was incorporated into the single grain disk model of IN05. Dust flung out from the inner regions of the disk is carried far above and below the mid-plane. This dust is directly exposed to stellar radiation so is hot enough to contribute to NIR flux, whilst also obscuring the direct stellar flux. This was shown to be a physically viable scenario for YSOs, including SU\,Aur, by \citet{Konigl11} and \citet{Petrov19}. The resulting model reveals a disk with an inner radius of $0.15$\,au and a dust-to-gas ratio of $0.01$. The inner radius is in agreement with the literature values discussed above. The flaring parameters $\alpha_{disk}$ and $\beta_{disk}$ were fixed such that $\alpha_{disk} = \beta_{disk}+1$ and found to be $2.4$ and $1.4$ respectively. The dusty disk wind mechanism can directly reproduce the flux ratio in the K-band, allowing for an good visibility fit and an improved SED fit. As the disk wind rises above the mid-plane it also shields the cooler parts of the disk, reducing the longer wavelength flux compared to rim-only models. This was compensated for in our models by increasing the scale height of the disk to $15$\,au at a radius of $100$\,au. However, the implementation of the dusty disk wind in this scenario is not completely physical, owing to the high into-wind outflow rate of $1\times10^{-7}\mathrm{ M_{\odot}yr^{-1}}$ required. The BK12 model also had the effect of flattening the second visibility lobe, a feature found in other YSOs \citep{Tannirkulam08,Setterholm18} and potentially opens powerful future modelling pathways for these objects. The grain size of the silicate dust species that produced the best fit was found to be $0.4\,\mathrm{\mu m}$ with no evidence of larger grains, as this addition resulted in a worse fit to the shape and magnitude of the NIR excess, particularly the shape of the Silicate feature around $10$\,um. This is in contrast to other inner disk studies where larger $1.2$\,um grains are required \citep{Kraus2009B,Davies18}.
    
    All the inner rim models investigated differ from the model proposed by \citet{Akeson05} whereby a vertical inner wall was combined with a small optically thick inner gas disk very close to the star aligned with the outer disk. This optically thick gas was implemented through very simple black-body emission models and allowed the author to successfully reproduced the observed NIR bump in excess flux. While TORUS could implement this type of black-body emission, it is unable to simulate the gas emission in a self-consistent physical manner.

\section{Conclusions}\label{Conc}
    This interferometric study of SU Aurigae has revealed the complex geometry and composition of the disk around SU\,Aurigae. We summarise our conclusions as follows:
    \begin{itemize}
        \item We reconstruct an interferometric image that confirms the inclined disk described in literature.  We see evidence for an asymmetry in the brightness distribution that can be explained by the exposure of the inner-rim on the far side of the disk and its obscuration on the near side due to inclination effects. Our data set does not permit the imaging fidelity that would be needed to detect evidence of ongoing planetary formation within the inner disk, such as small-scale asymmetries, gaps or rings.
        \item We see no evidence for a companion, in either the reconstructed images nor in the geometric model fitting procedures.
        \item Our simple geometric model fits reveal a disk of inclination $51.2\pm1.2^\circ$ along a minor axis position angle of $61.0\pm1.0^\circ$ and an inner radius of $1.12\pm0.12$\,mas ($=0.17\pm0.02$\,au). The disk is best modelled with a skewed ring which has a Gaussian ring width profile and sinusoidally modulated asymmetry. However, the poor $\chi^2_{red}$ of this model fit means the uncertainties quoted here are likely not representative of the the true range of values.
        \item Radiative transfer modelling shows that simple curved rim disk geometries of IN05 and THM07 cannot effectively model both the SED and visibility data. A deficit of NIR flux is obvious in the failure to reproduce our K-band observations. 
        \item A dusty disk wind scenario can successfully account for both the observed excess in the SED and the observed visibilities. The dusty disk wind scenario described here lifts material above the disk photosphere, thus exposing more dust grains to the higher temperatures close to the star responsible for the NIR excess. However, the high accretion rate required to reproduce the stellar-to-total flux ratio may make this scenario physically invalid.
        \item Our best-fit model (dusty disk wind model) suggests that the dust composition in the disk is dominated by medium sized grains ($0.4\,\mathrm{\mu m}$) with a sublimation temperature of $1600$\,K. Introducing larger grains results in a worse fit to the SED shape and NIR excess. The disk is also shown to be highly flared ($15$\,au at $100$\,au).
        \item The dusty disk wind model predicts a rather flat visibility profile at long baselines. This class of models avoids the pronounced visibility 'bounce' that are associated with sharp edges in brightness distributions, as predicted by rim-only models. Therefore, these models may also open a pathway to physically model other YSOs that have been observed with $\gtrsim\,300$\,m infrared long-baseline interferometry, such as AB\,Aur, MWC\,275, and V1295\,Aql \citep{Tannirkulam08,Setterholm18} which all observe very flat long baseline visibility profiles.
    \end{itemize}

\begin{acknowledgements}
    We acknowledge support from an STFC studentship (No.\ 630008203) and an European Research Council Starting Grant (Grant Agreement No.\ 639889).
    This research has made use of the VizieR catalogue access tool, CDS, Strasbourg, France. The original description of the VizieR service was published in \cite{Vizier}. 
\end{acknowledgements}

\bibliographystyle{aa}
\bibliography{REF}
\begin{table*}
    \centering
    \caption{\label{table:binaries} Binary fit parameters for the two calibrator stars HD\,31952 and HD\,34053.}
    \begin{tabular}{c c c c c c c}
        \hline
        \noalign{\smallskip}
        Star & Obs. Date & Flux Ratio & Separation [mas] & Position Angle [$^\circ$] & $\mathrm{UDD_{pri}}$ [mas] & $\mathrm{UDD_{sec}}$ [mas]  \\
        \hline
        \noalign{\smallskip}
        HD\,31952 & 2012-10-(19,20) & $3.76\pm0.07$ & $7.03\pm0.12$ & $95.55\pm2.35$ & $0.15\pm0.02$ & $0.15\pm0.03$\\
        HD\,34053 & 2012-10-(18,19,20) & $1.17\pm0.05$ & $1.66\pm0.23$ & $85.82\pm3.27$ & $0.14\pm0.02$ & $0.16\pm0.04$\\ [1ex]
        \hline
    \end{tabular}
    \tablefoot{
    Flux ratio is given as primary/secondary, the position angle is taken from North to East and UDD is the uniform disk diameter of the individual stars. 
    }
\end{table*}
\appendix
\section{Appendix A.}\label{AppA}
\subsection{Binary Fit Procedure}
The two calibrators HD\,31952 and HD\,34053 were found to be binary systems, based on strong non-zero closure phase signals. As both of these stars were needed for the calibration for 3 nights of our data, a binary fit was undertaken in order to recalculate the transfer function with which our data was calibrated. The software package \textit{LITpro} was used to construct the fits within a search radius of $10$\,mas.  The parameters fitted are: angular separation, position angle and the uniform disk diameters (UDD) of the primary and secondary components. Two uniform disks were used to represent the stars and $\chi^2$ maps were constructed to find the best-fit location of the secondary star. The parameters are displayed in Table~\ref{table:binaries}.

\section{Appendix B.}
Table of photometry used in the SED fitting procedure.
    \begin{table*}
        \centering
        \caption{\label{table:Photometry} Photometric values used to construct the SED of SU\,Aur}
        \begin{tabular}{c c c} 
            \hline
            Wavelength [$\mathrm{\mu m}$] & Flux [Jy] & Reference \\
            \hline
            0.15&1.31E-04&\citet{Bianchi11}\\
            0.23&0.00219&\citet{Bianchi11}\\
            0.42&0.233&\citet{Ammons06}\\
            0.44&0.402&\citet{Anderson12}\\
            0.53&0.6&\citet{Ammons06}\\\
            0.69&1.32&\citet{Morel78}\\
            0.79&1.47&\citet{Davies14}\\
            0.88&1.75&\citet{Morel78}\\
            1.24&2.08&\citet{Roser08}\\
            1.25&2.12&\citet{Ofek08}\\
            1.63&2.47&\citet{Ofek08}\\
            2.17&2.71&\citet{Roser08}\\
            2.19&2.62&\citet{Ofek08}\\
            3.35&2.6&\citet{Cutri14}\\
            3.40&2.44&\citet{Bourges14}\\
            4.50&1.75&\citet{Esplin14}\\
            4.60&2.78&\citet{Cutri14}\\
            5.03&2.58&\citet{Bourges14}\\
            7.88&1.99&\citet{Esplin14}\\
            8.62&2.36&\citet{Abrahamyan15}\\
            11.57&2.83&\citet{Cutri14}\\
            11.60&3.52&\citet{Abrahamyan15}\\
            18.40&6.47&\citet{Abrahamyan15}\\
            22.11&9.24&\citet{Cutri14}\\
            23.90&12.8&\citet{Abrahamyan15}\\
            61.89&12.2&\citet{Abrahamyan15}\\
            65.04&9.89&\citet{Toth14}\\
            90.06&8.8&\citet{Toth14}\\
            140.10&10.2&\citet{Toth14}\\
            160.11&8.88&\citet{Toth14}\\
            849.86&0.074&\citet{Mohanty13}\\
            887.57&0.071&\citet{Andrews13}\\
            1300.90&0.03&\citet{Mohanty13}\\
            1333.33&0.0274&\citet{Andrews13}\\ [1ex]
            \hline
        \end{tabular}
    \end{table*}

\end{document}